\documentclass[preprint,prd,a4paper]{revtex4}%
\usepackage{amsfonts}
\usepackage{amsmath}
\usepackage{amssymb}
\usepackage{graphicx}%
\providecommand{\U}[1]{\protect\rule{.1in}{.1in}}

\begin{document}
\preprint{gr-qc}
\title{The Friedmann Paradigm: A critical review}
\author{M.B. Altaie}
\affiliation{Department of Physics, Yarmouk University, 21163 Irbid, Jordan}
\keywords{Friedmann universe, singular universe, quantum universe.}
\pacs{PACS number}

\begin{abstract}
The Friedmann paradigm for a dynamical universe emanating from a spacetime
singularity is critically reviwed. Quantum effects, playing the essential role
at the very early stages, suggests that the universe may follow different
course to that presented by the standard Friedmann solutions. The
investigation of the back-reaction effect of the vacuum energy of quantized
massless matter fields at finite temperatures shows that the original spatial
singularity is avoided and that the universe is maintained all times at a
critical density. Instead of having a universe that was created at once we
have an emergent universe with energy being created continuously so as to
maintain the overall density at its critical value. The calculations presented
here provide a basis to construct a dynamical model for the universe where all
the known problems of the standard big bang can be avioded from start without
the need to assume the occurence of an inflation phase.

\end{abstract}
\volumeyear{year}
\date[Date text]{date}
\startpage{1}
\endpage{102}
\maketitle
\tableofcontents

\section{Introduction}

The Einstein field equations describes the relation between the geometry of
the spacetime and its matter/energy content by the equations%

\begin{equation}
R_{\mu\nu}-\frac{1}{2}g_{\mu\nu}R=-\frac{8\pi G}{c^{4}}T_{\mu\nu}, \label{q1}%
\end{equation}
where $R_{\mu\nu}$ are the components of the Ricci tensor, $R$ is the Ricci
scalar and $T_{\mu\nu}$ are the components of the energy-momentum tensor.

One more set of field equations were suggested by Einstein, which originally
were devised in order to freeze the dynamical universe described by
(\ref{q1}), these are%

\begin{equation}
R_{\mu\nu}-\frac{1}{2}g_{\mu\nu}R+g_{\mu\nu}\Lambda=-\frac{8\pi G}{c^{4}%
}T_{\mu\nu}, \label{q2}%
\end{equation}
where $\Lambda$ is the so-called cosmological constant. A positive $\Lambda$
represent a repulsive long- range force that may counter balance the
gravitational attraction between all parts of the universe. For this reason
and with an accurately chosen value of $\Lambda$\ Eqs. (\ref{q2}) can produce
a static universe. This was the original choice made by Einstein himself in
order to manipulate a static universe in accordance with the dominating
picture at that time. \ 

The construction of either (\ref{q1}) or (\ref{q2}) requires defining a basic
spacetime metric $ds^{2}$ in some general form and defining a matter/energy
distribution $T_{\mu\nu}$ to stand on the right-hand side. Consequently the
differential equations can be derived and solutions may be sought within
certain boundary conditions. The results enable us understand how such a model
universe will behave with time. Parameters calculated according to a given
model should be testable by real observations.

Confirmed astronomical observations performed during the second decade of the
last century suggested that the universe is expanding. This abandoned the
existence of a cosmological constant value that balances the gravitational
attraction, but did not necessarily abandon the presence of a parameter like
$\Lambda.$ This would allow for considering models of the universe with
non-zero $\Lambda$ which we will now call the "cosmological term". Such models
will be discussed later but first let us consider the more standard Friedmann
models for the universe.

\section{The Friedmann Solutions}

A homogeneous and isotropic distribution of matter/energy will lead to a
spherically symmetric spacetime. This can be best described by the
line-element ($c=1$)%

\begin{equation}
ds^{2}=dt^{2}-S^{2}(t)\left[  \frac{dr^{2}}{1-kr^{2}}+r^{2}\left(  d\theta
^{2}+\sin^{2}\theta d\varphi^{2}\right)  \right]  , \label{q10}%
\end{equation}

where $S(t)$ is a scale factor that describes the separation between any two
points on the spatial hypersurface%

\begin{equation}
d\sigma^{2}=\frac{dr^{2}}{1-kr^{2}}+r^{2}\left(  d\theta^{2}+\sin^{2}\theta
d\varphi^{2}\right)  . \label{q11}%
\end{equation}

In (\ref{q10}) the factor $k$ describes the curvature of the space; $k=+1$
describes positively curved space, whereas $k=0$ describes a flat space and
$k=-1$ describes the negatively curved hyperbolic space.

If we set%

\[
r=\sin\chi
\]

where $0\leq\chi\leq\pi,$ then with $k=1$ we have%

\[
d\sigma^{2}=d\chi^{2}+r^{2}\left(  d\theta^{2}+\sin^{2}\theta d\varphi
^{2}\right)
\]

This is a 3-dimentional spherical hypersurface ($S^{3}$) at which the
coordinates of any point are described by ($\chi,\theta,\varphi$). Therefore,
the spatial part of the metric in (\ref{q10}) is a 3-dimensional hypersurface
in 4-dimensional spacetime.

Alexander Friedmann adopted the metric in (\ref{q10}) and consequently
calculated the components of $R_{\mu\nu}$ as%
\begin{align*}
R_{0}^{0} &  =3\frac{\overset{\cdot\cdot}{S}}{S},\\
R_{1}^{1} &  =R_{2}^{2\text{ \ }}=R_{3}^{3}=\left[  \frac{\overset{\cdot\cdot
}{S}}{S}+\frac{2\overset{\cdot}{S^{2}}+2k}{S^{2}}\right]  ,\\
R &  ==6\left[  \frac{\overset{\cdot\cdot}{S}}{S}+\frac{\overset{\cdot}{S^{2}%
}+k}{S^{2}}\right]  ,
\end{align*}
He further assume that the energy momentum tensor is given by%

\begin{equation}
T_{\nu}^{\mu}=diag\left(  \rho,-p,-p,-p\right)  , \label{q12}%
\end{equation}
where $\rho$ is the energy density and $p$ is the pressure.

Using (\ref{q12}) the time-time component of (\ref{q1}) gives%

\begin{equation}
\left(  \frac{\overset{\cdot}{S}}{S}\right)  ^{2}+\frac{k}{S^{2}}=\frac{8\pi
G}{3}\rho\label{q13}%
\end{equation}

This is normally called the Friedmann equation. The spacial components yields
three identical equations, this is%

\begin{equation}
2\frac{\overset{\cdot\cdot}{S}}{S}+\left(  \frac{\overset{\cdot}{S}}%
{S}\right)  ^{2}+\frac{k}{S^{2}}=-8\pi G\text{ }p. \label{q14}%
\end{equation}

Subtracting (\ref{q13}) from (\ref{q14}) we get yet a third equation%

\begin{equation}
\frac{\overset{\cdot\cdot}{S}}{S}=-\frac{4\pi G}{3}\left(  \rho+3p\right)  .
\label{q15}%
\end{equation}

If we have to solve the above equations we need an equation which defines the
development of the matter/energy density, this could be obtained from the
so-called fluid equation which is essentially the law of conservation of
energy and momentum. The Friedmann paradigm assumes that%

\begin{equation}
T_{\nu;\mu}^{\mu}=0. \label{q16}%
\end{equation}

This gives the fluid equation%

\begin{equation}
\overset{\cdot}{\rho}+3\left(  \frac{\overset{\cdot}{S}}{S}\right)  \left(
\rho+p\right)  =0. \label{q17}%
\end{equation}

The condition in (\ref{q16}) also implies that the spacetime is conserved%

\begin{equation}
R_{\nu;\mu}^{\mu}-\frac{1}{2}g_{\nu}^{\mu}R_{;\mu}=0. \label{q18}%
\end{equation}

This conservation of the spacetime is carried through the varying
gravitational field within the spacetime while spacetime is developing, e.g.,
gravitation become weaker as the universe expand and becomes stronger as the
universe contract. Consequently the matter density has to vary like
$1/S^{3}(t).$ This normally is derived by solving the Friedmann equations for
a model of a pressureless dust universe.

In order to obtain the complete solutions of the Friedmann equation we need to
specify an equation of state which relates the momentum $p$ to the energy
density $\rho$, this is usually given by%

\begin{equation}
p=\omega\rho, \label{q19}%
\end{equation}

with $\omega=0,1/3,-1/3$ and $-1$ for pressureless dust, em radiation, vacuum
and cosmological term respectively.

Details of obtaining the different solutions of the Einstein field equations
according to the Friedmann paradigm can be found in \cite{Weinberg1},
\cite{Narlikar1} or \cite{barbara}.

\section{Parameters of the Friedmann solutions}

There are several basic parameters that can be designated for the Friedmann
solutions these are

\subsection{The Hubble parameter}

This is defined by%

\[
H=\frac{\overset{\cdot}{S(t)}}{S(t)}.
\]

At present time the value of $H$ is designated $H_{0}$ and is called the
\emph{Hubble constant}. The Hubble parameter defines the rate of the expansion
of the universe.

\subsection{The density parameter}

One important parameter that is deduced from the Friedmann solutions is the
density parameter given by%

\begin{equation}
\Omega=\frac{\rho}{\rho_{c}}, \label{q20}%
\end{equation}
where $\rho_{c}$ is a critical energy density defined by%

\[
\rho_{c}=\frac{3H^{2}}{8\pi G}.
\]

This parameter defines the geometry of the spacetime as being positively
curved when $\Omega>1,$ flat when $\Omega=1$ and is negatively curved when
$\Omega<1.$ The Friedmann equation can be re-written in terms of this
parameter as%

\begin{equation}
\Omega-1=\frac{k}{S^{2}(t)H^{2}}.\label{q21}%
\end{equation}

The density parameter do not change throughout the whole course of the
development of the universe. This implies that the state of curvature for the
universe stays as it is all the time, i.e., no change in the curvature mode.

\subsection{The deceleration parameter}

This is defined by%

\begin{equation}
q=-\frac{\overset{\cdot\cdot}{S(t)}}{S(t)}\frac{1}{H^{2}}, \label{q22}%
\end{equation}
the larger the value of $q$, the more rapid is the deceleration.

By adopting the condition (\ref{q16}) the standard model is assuming that all
the mass/energy content of the universe was capsulated in a \emph{singularity
},\emph{\ }from which the universe emanated. By solving the Friedmann equation
using the equation of state we conclude that the universe emanated rather
violently with acceleration at start. This caused to christen the model that
was based on the Friedmann paradigm as the \emph{big bang} model. This is why
it is always mentioned in the literature that the universe once was in the
state of a singularity with infinite density. However, such a claim is
physically unfounded.

\section{The Standard Big Bang Model}

During the late fourties of the last century a scenario for the production of
natural chemical elements were proposed by Gamow and collaborators. This
scenario considered a thermodynamical treatment of a Friedmann universe that
was assumed to be initially composed of a hot soup of elementary particles.
The scenario was able to explain the natural abundance of light elements only,
whereas other elements are found to be synthesized inside massive stars. The
standard big bang model predicted the existence of a relic cosmic microwave
background radiation (CMB) that was left over from the era when electrons were
combined with the nuclei to form atoms. This radiation is thought to be highly
homogenous and isotropic. Early calculations deduced that these radiation
should have the spectrum of\ a\ blackbody at temperature $T$ $\sim50$K but
later this figure was refined to be about $3K$.

In 1965 Penzias and Wilson discovered the cosmic microwave background
radiation. This discovery was considered to be the hard evidence for the
credibility of the big bang scenario, and therefore, for the underlaying
assumptions including those implied by the Friedmann solutions. This discovery
boosted the interest in the big bang model and research in cosmology.
Subsequently, this was grown into a sort of a paradigm, which I call the
\emph{Friedmann paradigm}. The whole content of the universe was thought to
have popped from nowhere at once with infinitely high temperature and went
expanding and cooling, breeding all sots of elementary particles that were
rushing all around in thermodynamical equilibrium. This paradigm dominated
modern cosmology during the last three decades and superseded all other
considerations like the steady-state theory of Hoyle and collaborators
\cite{Hoyle}. Refined measurements on a large angular scales confirmed the
main features of the CMB but marked higher accuracy in respect to the
homogeneity and isotropy.

Recent analysis of the CMB measurements indicates that the spatial part of the
universe is nearly flat. This requires that the average density of the
universe at the time of recombination be equal to the critical density. Since
no enough luminous matter is observed in the universe to cover the required
density, cosmologists assumed the existence of dark matter and (later dark
energy too) in order to compensate for the missing amount. In fact there is
nothing against such an assumption since our exploration of the universe is by
no means complete; it's only remain to find it.

The standard big bang model does not take into consideration any quantum
effects; such effects are thought to have played an important role at the very
early stages of the universe. As we will see later these effects rules-out the
possibility of the existence of a singularity and would define the later
course of the development of the universe.

\section{Friedmann Paradigm: A critical view}

The main points that feature the Friedmann cosmology are the followings:

\begin{enumerate}
\item The assumption that the total energy content of the universe is
conserved throughout the history of the universe allowing for violation only
at one initial moment.

\item The existence of an initial singularity in space and time.

\item The assumption that the universe is effectively one component closed
system with no outside; accordingly the total entropy is assumed to be
constant $dS=0$. This assumption implied that the expansion of the universe is
fully adiabatic.

\item The big question facing the Friedmann paradigm is how the universe
managed to cross its own horizon? A universe created according to the
Friedmann paradigm is apt to collapse under its own gravitational attraction.
It is not \ clear how the universe would go on expanding unless there is a
driving force within such a universe. Some important bit of physics is missing
here. The universe cannot work without a cosmological constant or vacuum energy.

\item No quantum effects were taken into consideration, for this reason the
homogeneity and isotropy of the spacetime was assumed to be perfect.
\end{enumerate}

The above features undermined the standard big bang model which was based on
the Friedmann paradigm and posed some serious problems like the horizon,
flatness and the formation of large cosmic structures and other problems. The
resolution of these problems needed the introduction of some sorts of the
quantum effects in the treatment of the very early universe. This was
introduced through an \textit{ad hoc} solution assuming the existence of
\emph{inflation}\ phase prior to the big bang phase. Technically this required
the assumption of a time-dependent scalar field that played the source for the
inflation \cite{Guth}. These suggestions were taken further and was developed
in to full fledge scheme later by Linde and others \cite{Linde1}, but
generally some basic criticism remains to be seriously valid unless a more
profound scheme is developed \cite{Liddle1}, \cite{Wald1} and \cite{Turock}.

\section{Quantum effects}

The consideration of the quantum effects requires the quantization of the
gravitational field. But since gravity, as best described by the theory of
general relativity, is non-linear therefore the standard canonical
quantization will not be suitable; the superposition principle is not
applicable and perturbation theory will not be consistent . For this reason we
have to resort to a semiclassical consideration in which matter fields are to
be considered quantized, whereas gravity is taken as a classical field. This
approximation has proved to be workable near and above the Planck scale
\cite{B&D}.

\section{Quantum Fields in Curved Spacetime}

The most interesting quantity to be considered for the quantum field
consideration in curved spacetime is the vacuum expectation value of the
energy-momentum tensor. The reason is that this quantity stands on the
right-hand side of the Einstein field equation and, therefore, would act as a
source for the energy in the universe. This is called the \emph{back-reaction
effect}. However, the direct calculations of the vacuum expectation value of
the energy-momentum tensor in a time-dependent closed universe, like the RW
universe, is cumbersome because of the difficulty in defining the vacuum state
in a time-dependent metric \cite{Ful}, and because of the anomalies that
appears in the trace of $T_{\mu\nu}$ \cite{B&D}. For these reasons the more
simpler case of the Einstein static universe were extensively considered and
it was found that $<0|T_{\mu}^{\nu}|0>$ in this universe is non-zero
\cite{Ford1}. By demanding self-consistency for the Einstein field equations,
this produces a non-singular universe \cite{Altaie1}. This encouraged us to
consider the finite temperature corrections to the vacuum expectation value of
several other field where it was shown that these corrections produces a
distribution of energy that, if utilized requiring self-consistent Einstein
field equations, would generate a temperature radius relationship that exhibit
some peculiar behavior \cite{Altaie2}, \cite{Altaie3} and \cite{Altaie4}. To
have a glance of the calculations here is a short summary.

The general structure of the energy density of the quantum fields at finite
temperatures is%

\begin{equation}
\mathtt{<}T_{00}\mathtt{>}_{tot}=\mathtt{<}T_{00}\mathtt{>}_{T}^{b}%
+\mathtt{<}T_{00}\mathtt{>}_{T}^{c}+\mathtt{<}T_{00}\mathtt{>}_{0},
\label{q23}%
\end{equation}
where $\mathtt{<}T_{00}\mathtt{>}_{T}^{b}$ is the flat space \textquotedblleft
black-body\textquotedblright\ term, $\mathtt{<}T_{00}\mathtt{>}_{T}^{c}$ is
the correction term arising from the finite size and the field-curvature
coupling, and $\mathtt{<}T_{00}\mathtt{>}_{0}$ is the zero-temperature vacuum
energy density.

In more compact form (\ref{q23}) can be written as%

\begin{equation}
\mathtt{<}T_{00}\mathtt{>}_{tot}=\mathtt{<}T_{00}\mathtt{>}_{T}+\mathtt{<}%
T_{00}\mathtt{>}_{0}, \label{q24}%
\end{equation}
\hspace{5mm} \hspace{5mm} \hspace{5mm} \hspace{5mm} \hspace{5mm} \hspace{5mm} where%

\begin{equation}
\text{\texttt{%
$<$%
}}T_{00}\text{\texttt{%
$>$%
}}_{T}=\frac{1}{V}\sum\limits_{n}\frac{d_{n}\epsilon_{n}}{\exp\beta
\epsilon_{n}\pm1}, \label{q25}%
\end{equation}
in which $\beta$ $=1/T$ , ${\epsilon}_{n}$ and $d_{n}$ are the energy
eigenvalues and degeneracy of the $n^{\text{th}}$ state respectively. The plus
sign is for bosons and the minus sign is for fermions. Throughout the rest
this paper we will use the system of units in which $c=G=\hbar=k=1$.

The back reaction effect is then calculated by substituting $\mathtt{<}%
T_{00}\mathtt{>}_{tot}$ on the right hand side of (\ref{q2}) and require a
self-consistent solution for the Einstein field equation. Because of the
symmetry properties enjoyed by the Einstein universe the general solution
always results in a simple form%

\begin{equation}
\frac{3}{a^{2}}=16\pi<T_{00}>_{tot}. \label{q26}%
\end{equation}

\subsection{The conformaly coupled scalar field}

The conformaly coupled massless scalar field satisfies the equation%

\begin{equation}
\nabla_{\mu}\nabla^{\mu}\Phi-\frac{R}{6}\Phi=0, \label{q27}%
\end{equation}
where $\nabla_{\mu}$ is the covariant derivative and $R$ is the scalar curvature.

The expression for the $T-a$\textit{\ }$\ $relationship for the conformaly
coupled massless scalar field as deduced from the calculations of the back
reaction effects is given by%

\begin{equation}
a^{2}=\frac{8}{3\pi}\sum\limits_{n=1}^{\infty}\frac{n^{3}}{e^{n/Ta}-1}%
+\frac{1}{90\pi}.\; \label{q28}%
\end{equation}
\hspace{5mm} \hspace{5mm} \hspace{5mm} \hspace{5mm} \hspace{5mm}

As is shown in Fig. 1., this relationship exhibit a minimum radius $a_{0}$ $=
$ $(1/90{\pi})^{1/2}l_{p}$ at $T=0$, and the transition temperature is
$T_{max}$ $=2.218T_{p}$, at which the radius of the universe is given by
$a_{c}=0.072l_{p}$.

\subsection{The minimally coupled massless scalar field}

The minimally coupled massless scalar field satisfies the equation%

\begin{equation}
\nabla_{\mu}\nabla^{\mu}\phi=0. \label{q29}%
\end{equation}

The vacuum energy density at $T=0$ for this field is zero. Therefore, the
result of the back reaction of the field at finite temperatures will contain
the contributions from the energy mode-sum only. Accordingly, the $T-a$
relationship is given by%

\begin{equation}
a^{2}=\frac{8}{3\pi}\sum\limits_{n=0}^{\infty}\frac{(n+1)^{2}[n(n+2)]^{1/2}%
}{e^{[n(n+2)]^{1/2}/Ta}-1}. \label{q30}%
\end{equation}

Results of the back reaction study shows that the universe will be singular in
presence of this field. The transition temperature is $T_{max}$ $=0.6T_{p} $,
occurring at a radius $a_{c}=0.68l_{p}$.

\subsection{The neutrino field}

The field equation is given by%

\begin{equation}
\gamma_{\mu}\nabla^{\mu}\psi=0, \label{q31}%
\end{equation}

where $\gamma_{\mu}$ are the Dirac matrices.

The $T-a$ relationship is then given by%

\begin{equation}
a^{2}=\frac{16}{3\pi}\sum\limits_{n=1}^{\infty}\frac{n(n+1/2)(n+1)}%
{e^{(n+1/2)/Ta}+1}+\frac{17}{180\pi}. \label{q32}%
\end{equation}

The results obtained from the study of back reaction indicates that this
relation exhibit a minimum radius $a_{0}=(17/180{\pi})^{1/2}$ $l_{p}$, and
transition temperature $T_{max}=1.076T_{p}$ at a critical radius
$a_{c}=0.204l_{p}$.

\subsection{The Photon field}

The free photon field satisfies the covariant equation%

\begin{equation}
\nabla_{\mu}F^{\mu\nu}=0. \label{q33}%
\end{equation}

The study of the back reaction effects resulted in the following $T-a$ \ relationship%

\begin{equation}
a^{2}=\frac{16}{3\pi}\sum\limits_{n=2}^{\infty}\frac{n(n^{2}-1)}{e^{n/Ta}%
-1}+\frac{11}{45\pi}. \label{q34}%
\end{equation}

This relationship exhibits a minimum radius $a_{0}=(11/45{\pi})^{1/2}l_{p}$,
and a transition temperature $T_{max}$ $=1.015T_{p}$ taking place at
$a_{c}=0.34l_{p}$.

The results of these calculations are depicted in FIG. 1. Generally we notice
that with small radii around the Planck scale the temperature rises
exponentially with the radius, whereas at large radii the energy of the
universe behaves according to the Planck blackbody radiation law. An
interesting feature of this temperature-radius relationship is that it exhibit
a maximum temperature at the Planck scale. This maximum separates between what
we call the \emph{Casimir regime} \ and the \emph{Planck regime} \ and
indicates a change of the physical state of models with radii above certain values.

To explain the transition from the Casimir regime into the Planck regime and
because this is a sort of a phase transition taking place at a very early
stage of the universe, we considered the phenomena of Bose-Einstein
condensation in an Einstein universe. Consideration of the Bose-Einstein
condensation of spin 1 field in the ultra-relativistic limit
\cite{Altaie&Malk} showed that the critical condensation temperature is the
maximum temperature itself obtained in considering the back-reaction effect.
This is a remarkable result that would certainly need further investigation.

\section{The cosmological constant}

As I mentioned earlier, the cosmological constant was first introduced by
Einstein in order to construct a static universe that will not collapse
against its own gravitational attraction. The discovery of Hubble that the
universe may be expanding led Einstein to abandon the idea of a static
universe and, along with it the cosmological constant. Recent year have
witnessed a resurgence of interest in the possibility that a positive
cosmological constant $\Lambda$ may dominate the total energy density in the
universe (for recent reviews see \cite{carrol} and \cite{Sahni}). At a
theoretical level $\Lambda$ is predicted to arise out of the zero-point
quantum vacuum fluctuations of the fundamental quantum fields. Using
parameters arising in the electroweak theory results in a value of the vacuum
energy density $\rho_{vac}=10^{6}$ GeV$^{4}$ which is almost $10^{53}$ times
larger than the current observational upper limit on $\Lambda$ which is about
$10^{-47}$. This means that GeV$^{4}\sim10^{-29}$ gm/cm$^{3}$. On the other
hand, the QCD vacuum is expected to generate a cosmological constant of the
order of $10^{-3}$ GeV$^{4}$ which is many orders of magnitude larger than the
observed value. This is known as the old cosmological constant problem
\cite{Weinberg2}. The new cosmological problem is to understand why
$\rho_{vac}$ is not only small but also, as the current observations seem to
indicate, is of the same order of magnitude as the present mass density of the universe.

The value of the cosmological constant for an Einstein universe seem to be
trivial. It is directly related with the total energy density. However, since
the energy density in an Einstein universe varies inversely with $a^{2}$ and
not with $a^{3}$, new features are expected in the behavior of the
cosmological constant. In what follows we are going to investigate the
possible values of the cosmological constant for different radii of the
Einstein universe in presence of the massless matter fields. But since
different radii of the universe corresponds to different temperature with a
non-trivial relationship between the radius and the temperature as was found
earlier, the values of the cosmological constant at different temperatures
turns out to be non-trivial and is rather of some serious interest as we find
a qualitative differences between the conformaly coupled and minimally coupled
scalar fields.

Field theorists and particle physicists insist on the value they obtain for
the cosmological constant \cite{Weinberg2}. However this is much larger than
that obtained by analyzing recent measurements of CMB \cite{Bennet}. The
reason is that they cannot see a resolution for this discrepancy and have no
alternative to their standard model of particle physics. But from dimensional
argument $\Lambda$ should be small now. One major point which particle
physicists seem to have neglected is the difference between conformaly coupled
scalar field and the minimally coupled scalar field. Despite the fact that
both fields are alike in the present universe, it may be of some importance to
know that this difference is essential in the very early universe. In fact, a
minimally coupled field do not exist in nature unless the universe is
absolutely flat \cite{Altaie&Malk}.

The results of these studies show that the Einstein "toy" model considered
here can explain the low present-value of the cosmological constant
\cite{Altaie3}. It is found that the cosmological constant takes nearly the
same value for small radii and then at certain radius (the value where maximum
temperature occurs) solutions shows that the value of the cosmological
constant decays quickly for smaller and smaller values.

Contracting Eq. (\ref{q2}) and taking into consideration that $R_{00}=0$ in
the static Einstein universe and that $T_{%
\mu
}^{\mu}$ vanishes for massless fields, we obtain%

\begin{equation}
\Lambda=\frac{R}{4}=\frac{3}{2a^{2}}. \label{q35}%
\end{equation}

On the other hand, for the case of Einstein static universe, the field
equation reduces to%

\begin{equation}
-\frac{3}{a^{2}}+\Lambda=-8\pi\rho_{tot}., \label{q36}%
\end{equation}

and%

\begin{equation}
-\frac{1}{a^{2}}+\Lambda=\frac{8\pi\rho_{tot}}{3}, \label{q37}%
\end{equation}

where%
\begin{equation}
\rho_{tot}=<0|T_{0}^{0}|0>. \label{q38}%
\end{equation}

Here we will consider $\rho_{tot}=\rho_{vac}+\rho_{rad},$ but in a more
general case one can set $\rho_{tot}=\rho_{vac}+\rho_{rad}+\rho_{matter}$,
with $\rho_{rad}$ belonging to the massless field filling the spatial part of
the universe and $\rho_{matter}$ belonging to the pressureless dust that may
exist. The addition of the energy density of the pressureless matter will not
make any qualitative change in the results since $\rho_{matter}$ in an
Einstein universe specifically behaves same as $\rho_{vac}$ and $\rho_{rad}.$

Solving (\ref{q36}) and (\ref{q37}) we obtain%

\begin{equation}
\Lambda=8\pi\rho_{tot}. \label{q39}%
\end{equation}

Using and the results obtained in the previous section we obtain the results
depicted in FIG. 2. These results show that the conformaly coupled scalar, the
neutrino and the photon field have similar behaviors. For these three fields
the value of the cosmological constant during the Casimir regime ( the vacuum
era) is high and is nearly constant, a point which is required by the
inflationary models. Also, it is very important to notice that the
cosmological constant decays to very small values when the size of the
universe is already within the Planck scale. This behavior comes in agreement
with what inflation theories is suggesting.

Particle physicist find that the vacuum energy is given by
\begin{equation}
\varepsilon_{vac}\sim\frac{E_{P}}{l_{P}^{3}}, \label{40}%
\end{equation}
and they think that this value should be constant and independent of the
radius of the universe (see for example \cite{barbara}, p.60). But in fact
studies of quantum fields in curved spacetime indicates that this belief is
not true. In the Einstein universe, at the least, the vacuum energy is
directly related to the radius of the universe. There is no reason why should
the same would not apply to the time-dependent RW universe. Particle
physicists may need to revisit their theory for some fundamental
considerations. For example the scalar field by which typical calculations are
carried in the standard model of particle physics is usually taken to be
minimally coupled, i.e., the curvature coupling is considered to be zero.
However our considerations of quantum fields in curved spacetime
\cite{Altaie&Malk} shows that the minimally coupled scalar field may not be
the proper one to consider, especially at the early stages of the universe for
the essential effect of the curvature term which acts as a correction to the
mass term. Conformal invariance should be taken seriously in seeking proper solutions.

\section{The value of Einstein Universe}

The above calculations and results did not seem to attract the attention of
physicists working to solve problems with contemporary cosmology. The reason
is: the Einstein universe is static, and no one would expect that it would
suggest anything of realistic value. Physicist expect that a realistic model
for the universe must be dynamical taking into consideration the reality of
the expansion of the universe. Moreover, and in analogy with electrodynamics,
physicist insist on a dynamical consideration since they expect that some
gravitational effects, like particle creation, will be produced consequently,
which will certainly back-react on the whole model.

In fact there are enough reasons to support our belief that studies of some
physical parameters in an Einstein universe can provide a picture of the
development of the universe at very early stages. Such results will,
qualitatively at the least, be correct. However the Einstein static universe
remained to be of interest to theoreticians since it provided a useful model
to achieve better understanding of the interplay of spacetime curvature and of
quantum field theoretic effects.

The conformal relationship between the static Einstein universe and the
Robertson-Walker universe and the possibility to consider the Einstein
universe of a given radius as representative of an instantaneously static
Robertson-Walker universe \cite{Ford1} and the ($1-1$) correspondence between
the vacuum and the many particle states of both universes as established by
the work of Kennedy \cite{Ken}, suggests that the thermal behavior of a real
closed universe is qualitatively similar to the results obtained in this work.
Therefore, we feel that the calculations in the Einstein universe are useful
in understanding the interplay between quantum fields and the curvature.
Indeed our calculations showed that an Einstein universe with curvature radius
of about two order of magnitude larger than the Hubble radius will have the
same CMB temperature as the presently measured one. On the other hand the
analysis of the most recent observations of CMB spectrum suggests that the
curvature radius of the real universe is at least $5$ times larger than the
Hubble radius \cite{Turner}. This is a point in favour of the practical
relevance of our calculations.

Despite the fact that the Einstein universe is static we can imagine a series
of successive states of Einstein universe developing according to parameters
that are related self-consistently according to the Einstein field equations.
In this context a dynamically developing universe may be imagined as a series
of successive static states each described by the Einstein static model. This
obviously is related to the philosophical question of whether infinite
divisibility can ontologically exist at all. In such a model the
geometro-dynamical effects like particle production by changing geometry is
obviously lost. However, because general relativity is a self-consistent
theory, therefore in effect such a model is expected to exhibit a consistent
overall behavior so that the end result will not be different from those
obtained in a fully dynamical model. We claim that this property is specific
to the Einstein universe which is conformal to the RWF universe. Indeed, it is
the property of the Einstein universe of having matter content without motion,
but then the geometry of the universe is strictly related to its matter
content such that any larger universe would necessitate an increase in the
material content of the universe. In a dynamically analogous model this means that%

\[
\overset{\cdot}{M}=\overset{\cdot}{a},
\]
which may be taken to compensate for the particle creation in the dynamical
state.\bigskip

I feel that the results presented in this paper, are quite encouraging to
construct a new dynamical model that starts from Planck parameters evolving to
the present stage with the total density falling like $1/a^{2}$ rather than
$1/a^{3}$. However, such a model will involve continuous particle creation at
a rate proportional to Hubble constant. Obviously a mechanism for the particle
creations has to be devised for such a process. In this respect the mechanism
suggested by Hoyle-Narlikar (\cite{HN} and \cite{Narlikar}) in the context of
the steady-state theory may be useful, however the rate of creation will be
different from that of the steady state theory.

Furthermore we note that our calculation shows that a universe stemming from
the vacuum and developing by the availability of the vacuum energy would be
maintained to be at the critical density. Accordingly a justification can be
provided be construct a non-singular universe free from the problems of the
standard big bang model. We mean that our calculations will provide the
necessary justifications for the consideration of Ozer and Taha \cite{Ozer} of
a critical density universe. On the other hand this will explain why do we
have a flat or nearly flat universe without the need for inflation. However it
should be pointed out here that what remains is the remarkable success of the
standard big bang scenario ( which utilizes the Friedmann paradigm) in
explaining the natural abundance of light elements. If any viable alternative
to the Friedmann paradigm is to be presented then it should be able to explain
the natural abundance of light elements (see \cite{Abd}).\bigskip

\textbf{Figure Caption:}

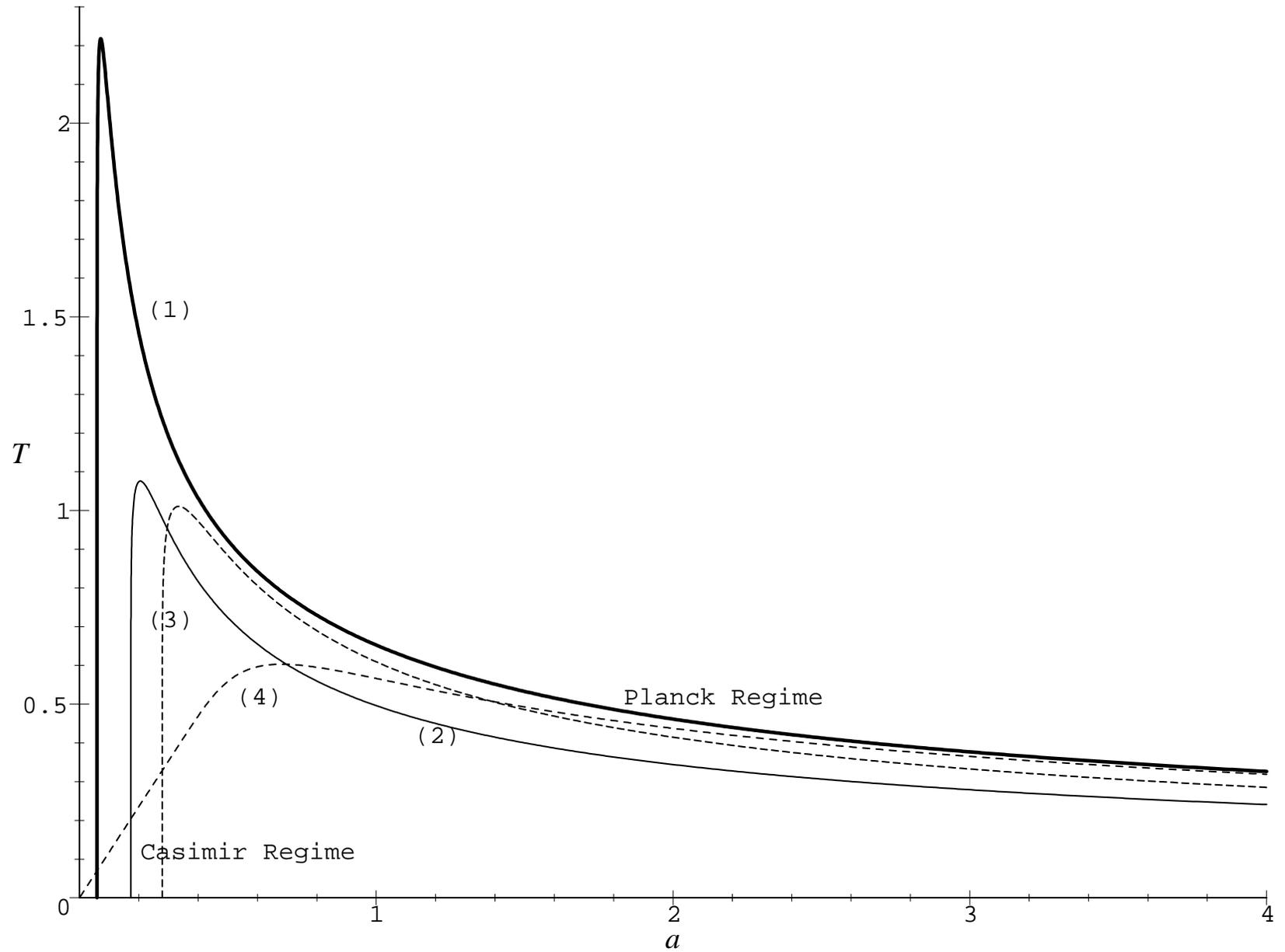
\textbf{FIG. 1}: The temperature-radius relationship deduced from the finite
temperaure corrections to the vacuum energy ploted for different matter
fields: the conformally coupled scalar field (1), the neutrino field (2), the
photon field (3) and the minimally coupled field (4). (see Ref. \cite{Altaie5}).

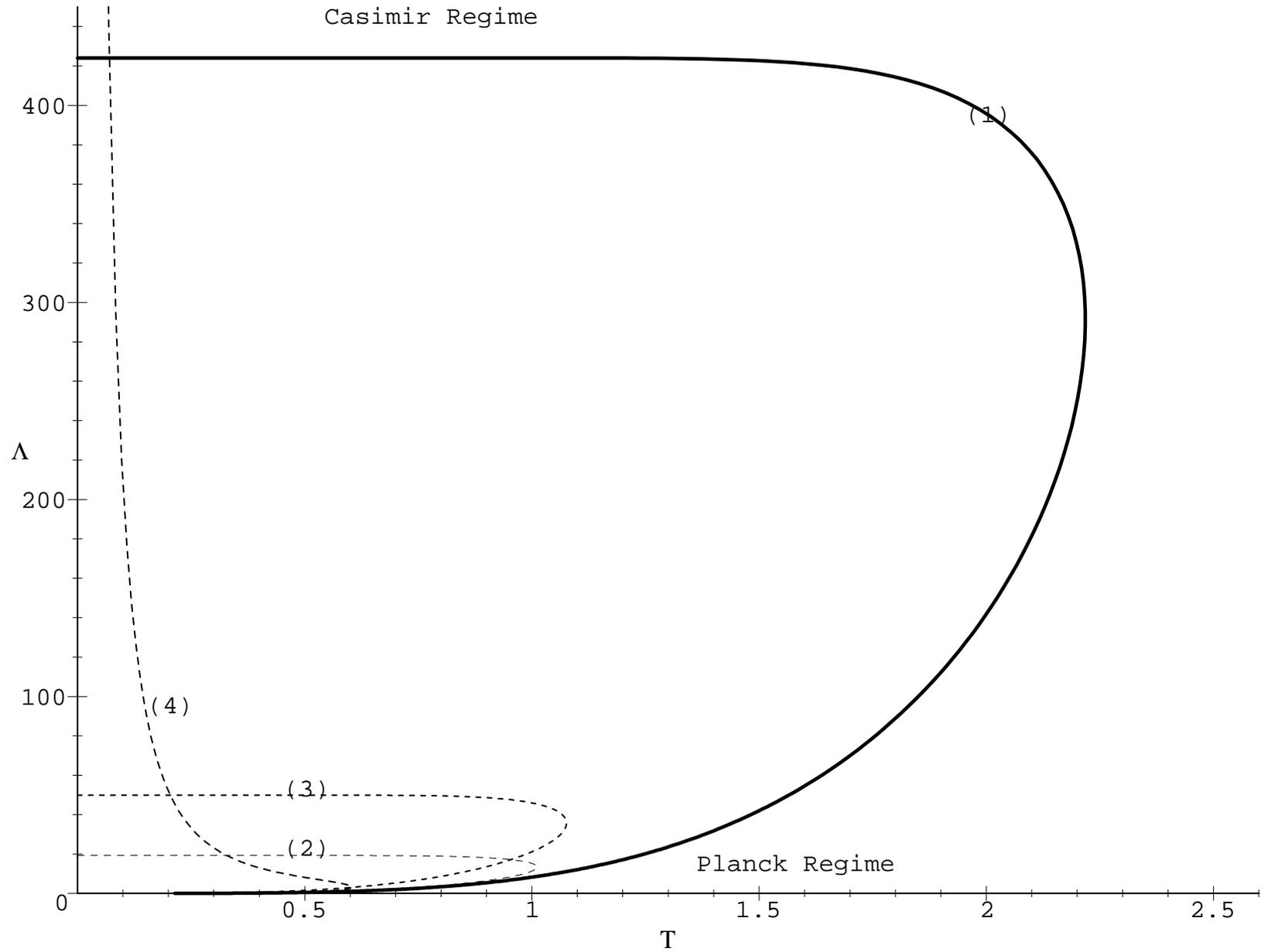
\textbf{FIG. 2}: contributions of the conformally coupled scalar field (1),
the photon field (2), the neutrino field (3) and the minimally coupled scalar
field (4) to the cosmological constant in an Einstein universe at finite
temperatures. (see Ref. \cite{Altaie5}).

\end{document}